%% file: main.tex
\documentclass{article}
\usepackage{amsmath,amsfonts,graphicx}
\usepackage{color}
\usepackage{caption}
\usepackage{subcaption}
\usepackage{siunitx}
\usepackage[margin=1.0in]{geometry}
\usepackage{url}
\usepackage{hyperref}
\usepackage[dvipsnames]{xcolor}
\hypersetup{
    colorlinks=true,
    linkcolor=blue,
    citecolor=BlueViolet,
    filecolor=magenta,      
    urlcolor=RoyalBlue,
    pdftitle={Android Private Compute Core Architecture},
    pdfpagemode=FullScreen,
    }
\usepackage{fancyhdr}
\usepackage{multirow}
\usepackage{array}
\usepackage{tabularx}
\usepackage{textcomp}

\usepackage[
sorting=none
]{biblatex}
\title{Private Compute Core
Architecture}
\author{Eugenio Marchiori, Sarah de Haas, Sergey Volnov,\\
Ronnie Falcon, Roxanne Pinto, and Marco Zamarato}
\date{Google\\
\vspace{1em}
    \today
}

\begin{document}
\maketitle

%

%

\input{01_introduction}
\input{02_system_architecture}
\input{03_Implementation_examples}

\input{04_conclusion.tex}

\clearpage
\printbibliography

\medskip
\end{document}

%% file: 01_introduction.tex
\section{Introduction}
Android’s Private Compute Core (PCC) is a secure, isolated environment within the operating system, that maintains separation from apps while enabling users and developers to maintain control over their data \cite{1}. It is backed by open-source code in the Android Framework introduced in Android 12 \cite{2}. PCC allows features to communicate with a server to receive model updates and contribute to global model training through Private Compute Services (PCS), the core of which has been open sourced\footnote{Details of what has been open sourced can be reviewed at Private Compute Services. Google Open Source Project on GitHub. Retrieved Jul 2022 from \url{https://github.com/google/private-compute-services}.}.\\

PCC is part of the OS, and by virtue of being isolated, constrained, and trusted, it can host sophisticated ML features. The hosted features themselves, running inside PCC, can be closed source and updatable.  In this way, PCC enables machine learning features to process ambient and OS-level data and improve over time, while restricting the availability of information about individual users to servers or apps.\\

With each new Android release we’ll add more features to the Private Compute Core. Some examples of features that now operate within PCC include:

\begin{itemize}
    \item \href{https://blog.google/products/android/live-caption/}{Live Caption}, which adds captions to any media using Google’s on-device speech recognition 
    \item \href{https://support.google.com/pixelphone/answer/7535326}{Now Playing}, which recognizes music playing nearby and displays the song title and artist name on the device’s lock screen
    \item \href{https://www.android.com/android-11}{Smart Reply}, which suggests relevant responses based on the conversation you’re having in messaging apps
\end{itemize}

%% file: 02_system_architecture.tex
\section{System architecture}
\subsection{Data handled/managed by PCC}
In this document, “ambient and OS-level data” refers to three main categories:

\begin{itemize}
\item \textbf{Raw data:} this is usually from the device sensors (camera, microphone, etc) or generated by the OS for its normal operation (such as screen content). This data is shared by the operating system with PCC so that it can manage feature access and its usage. This is referred to as “ambient and OS-level data” or just "data" throughout this document.
\item \textbf{Derived data:} this is data generated from analysis or inferences based on ambient and OS-level data. Derived data can in some cases be allowed to leave the PCC environment, subject to technically enforced policy.
\item \textbf{Metadata from system operation} this data is considered non-user identifiable if the user retains k-anonymity\cite{3} (where k is feature and context dependent), but some can be identifiable and thus subject to stricter precautionary measures.
\end{itemize}

The following is a non-exhaustive categorization of collected ambient and OS-level data in PCC:
\begin{table}[ht]
    \centering
    \small
    \setlength{\tabcolsep}{6pt}
\renewcommand{\arraystretch}{1.5}
\begin{minipage}{\textwidth}
\begin{tabular}
[c]{|l|l|l|l|}
\hline
\scshape Category & \scshape Type & \scshape Source & \scshape Method of access \\
\hline
\multirow{7}{4em}{OS-level} & audio & framework audio (e.g. song currently playing) & via audio APIs\\
\cline{2-4}
& image & screenshots (i.e. when changing activity) & via Content Suggestions API\\
\cline{2-4}
\multirow {2}{4em} & {text} & screen capture & via Content Capture API\\
\cline{3-4}
& & notification content & via NLS\footnote{NotificationListenerService} and NAS\footnote{NotificationAssistantService}\\
\cline{2-4}
\multirow {4}{4em} & {structured} & app data (indexed) & via AppSearch API\\
\cline{3-4}
& & app data (pushed) & via ContentCapture API\\
\cline{3-4}
& & contacts & via Contacts API \\
\cline{3-4}
& & app launches & via UsageStatsManager \\ 
\cline{3-4}
& & shortcuts & via ShortcutManager\\
\hline
\multirow{3}{4em}{ambient} & audio & background microphone & via audio APIs\\ 
\cline{2-4}
& image & background camera & via camera APIs \\ 
\cline{2-4}
& location & GPS and other signals & via Location Provider \\ 
\hline
\end{tabular} 
\end{minipage}
\normalsize
\caption{Types and sources of data used in PCC}
\end{table}

\newpage
\subsection{Overview diagram}
The current high-level architecture of Private Compute Core is pictured in Figure 1. PCC exists in its own isolated virtual sandbox (following a verifiably private design), with features allowed to run computations inside that sandbox on ambient or OS-level data generated by the user. The results are surfaced to the user via the UI operated by the trusted OS, or through access-controlled (by permissions or other mechanisms) open-source framework APIs as needed\footnote{Minimal logging uses Android Framework APIs and is limited to non-identifiable data such as performance and stability metrics.}. \\

The inferences from ambient and OS-level data are retained inside Android System Intelligence, within Private Compute Core. All communications to external servers is done via Private Compute Services (PCS) privacy-preserving open-source technologies. 

\begin{figure}[ht!]
\centering
\includegraphics[width=0.95\textwidth]{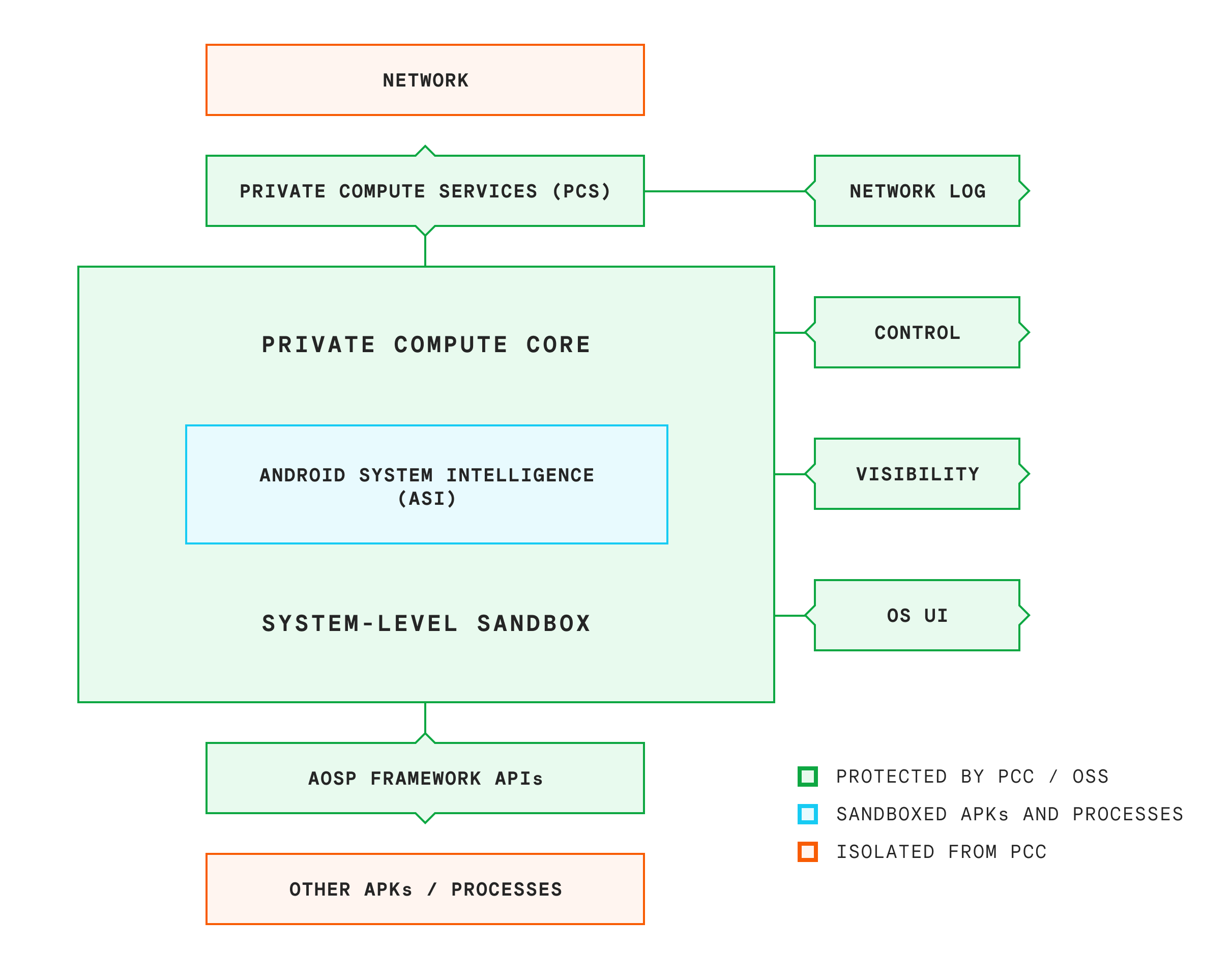}
\caption{Conceptual architecture diagram of Private Compute Core}
\end{figure}

\subsubsection{Android System Intelligence}
\href{https://play.google.com/store/apps/details?id=com.google.android.as}{Android System Intelligence (ASI)} is a Google-proprietary package containing system-level features like Live Caption\cite{4}, Now Playing\cite{5}, and others. It can only be accessed on compatible devices, and the available features may vary.\\

ASI restricts usage of the data through mechanisms like the \verb|<allow-association>| system (Table 2), or the absence of \verb|android.permission.INTERNET|\footnote{\url{https://developer.android.com/reference/android/Manifest.permission\#INTERNET}} permission (through the system permission system). It solely relies on PCS (described in detail below) and the Android framework for any information egress, both of which are open-source. 

\subsubsection{Private Compute Services}
Starting in Android 12, features within PCC no longer have \texttt{android.permission.INTERNET} permission, to allow claims about what data is shared with the network to be externally verifiable. To retain functionality without internet capability and increase the transparency of how features use network access, \href{https://security.googleblog.com/2021/09/introducing-androids-private-compute.html}{Private Compute Services (PCS)} was created. \\

\href{https://play.google.com/store/apps/details?id=com.google.android.as.oss. }{Private Compute Services (PCS)} is an APK with open-source\footnote{PCS code is continuously published to \url{https://github.com/google/private-compute-services/}}, code, which provides a narrow set of structured APIs purposefully designed for PCC components. Currently, PCS supports:

\begin{enumerate}
\item \textbf{Federated learning\cite{7} and federated analytics\cite{8},
} enabling privacy-preserving machine learning and analytics without centralized data collection. The underlying techniques involve pushing a computation graph (e.g. machine learning model) to the device, computing on the locally stored data, and sending only the computation results back. The results from many devices are aggregated together, and used to improve the device features and user experience. Each individual device’s results are protected from being seen by the orchestrating server through the use of the Secure Aggregation\cite{9} multi-party computation protocol, ensuring that only aggregates over many (e.g. thousands) of devices are made available to servers and model/feature developers.
\item \textbf{Private Information Retrieval (PIR)\cite{10}}, enabling privacy-preserving lookups of information from the cloud. The PIR protocol relies on a client-side implementation of homomorphic encryption to retrieve resources from a server-hosted database, while cryptographically ensuring that the server is unable to observe which resource was downloaded. 
\item \textbf{An HTTPS download-only transport,} which limits downloads to pre-specified global ML models. 
\end{enumerate}

\subsection{Isolation mechanisms}

\subsubsection{IPC bind restrictions}
PCC sets constraints on data sharing both for off-device services and those working on-device. On the client, technical limits are set on the other processes in the operating system (including apps) that PCC features can communicate with.\\

This is called the Inter Process Communication (IPC) allow-association mechanism, and it defines the exhaustive list (allow-list) of packages that a PCC package can establish an IPC channel with (outside the existing Android API surface), or vice versa. The recommended list of packages in the allow-list is defined in the \href{https://source.android.com/compatibility/12/android-12-cdd}{CDD (Compatibility Definition Document)} under ContentCapture and AppSearch.\\

As a result, only the system can establish communications between PCC and other services, which allows for inspectable API calls with any process not in the allow-list.

\subsubsection{Process isolation}
Some instantiations of PCC are not applicable at the package level, but instead only affect a single feature. For these, the system can use Android’s isolated processes\footnote{\url{https://developer.android.com/guide/topics/manifest/service-element\#isolated}}\cite{11}, in conjunction with other means (e.g. permission management), which are designed to ensure that the data remains separate from the application shipping the code and all APIs are open-source and inspectable.\\

Through this mechanism, it's possible to build features which follow the PCC implementation pattern, but whose code might be provided by an APK that fulfills other roles. Some versions of this also force these processes to be ephemeral (starting new ones regularly to replace the old ones).

\subsection{Permissions and controls}
\subsubsection{Android permissions}
Data going in and out of PCC features does so via Android Framework APIs. By extension, this means that such data flows are governed by the Android permission framework\footnote{\url{https://developer.android.com/guide/topics/permissions/overview}}. Given that most sensitive permissions can only be pre-granted by the OS (via explicit allow lists or roles), this can’t be modified after the fact, except on an operating system update. Anyone is able to analyze requested permissions by PCC components on a device at install time (by looking at the manifest) and runtime (via Android Debug Bridge\footnote{\url{https://developer.android.com/studio/command-line/adb}}).

\subsubsection{Android platform controls}
PCC features depend on APIs defined in the platform. Other implementations of features in the Android ecosystem that use screen content or app search data are subject to these controls, as they are required to pass \href{https://source.android.com/compatibility/cts}{CTS (Compatibility Test Suite)} and comply with Android’s \href{https://source.android.com/compatibility/cdd}{CDD (Compatibility Definition Document)}. \\

Android System Intelligence functions as a non-replaceable system component. Although it has its own separate update process, Android will check signatures upon each update making sure that it cannot be easily replaced by implementations from a different source. 

\subsubsection{User controls}
PCC features allow data to leave the designated sandboxes under limited circumstances. When possible the controls are implemented as part of the platform and not managed by the PCC feature itself.\\

PCC implementations offer additional ways for users to control if and how features use their data, and management of the data retained by those features. These include:
\begin{enumerate}
\item In Android OS, “Settings\textrightarrow Privacy\textrightarrow Personalize using app data”. This setting is part of AOSP and restricts screen data, app search data, and most other app-linked data from being shared with PCC features.
\item “Clear data” controls, offered by the features themselves. These are usually not open-source, but allow for granular control in many cases, including deleting all data or only for a period of time.
\item Android runtime permissions, controlled by AOSP, which the user can revoke. 
\item  Android sensor toggles, which allow the user to disable the camera or microphone, and prevent the data from these sensors from being used across the OS including by PCC features.
\end{enumerate}

\subsubsection{App developer controls}
Following the Android Platform Security Model\cite{11} developers maintain control of data generated by their own applications.\\

An example of a control is provided by \texttt{FLAG\_SECURE}\footnote{\url{https://developer.android.com/reference/android/view/WindowManager.LayoutParams\#FLAG\_SECURE}}. This is a mechanism, defined in Android, that allows apps to opt-out of operating system functionalities (in this case screenshots or being viewed on non-secure displays). By extension, this also applies to data processing in PCC from activities included under this setting. In some cases, more granular controls are provided on a per-feature basis.

\subsection{Surfaces}
PCC features might interact with surfaces that are outside of the PCC environment. To achieve this, we define two types of surfaces - Public Framework APIs and Operating System User Interfaces (which also use APIs): 

\subsubsection{Public Framework APIs}
In some cases we expose PCC features through Framework APIs (for example, a speech recognizer\footnote{\url{https://developer.android.com/reference/android/speech/SpeechRecognizer.html\#createOnDeviceSpeechRecognizer(android.content.Context)}} which ensures speech data is protected). For cases like these, we implement the following: 
\begin{enumerate}
\item \textit{Framework proxies: }to make sure that feature updates still follow the same rules, all API communications with PCC features are proxied by the Android platform. This means that the platform can enforce rules on how they are used (e.g. rate limiting) and what data is passed on to other apps (e.g. no side channels).
\item \emph{Data limitations:} data can only be used if it will not result in user identification or escalation of permission by the caller app. 
\end{enumerate}

\subsubsection{Operating System User Interfaces}
As a general rule, exposing PCC data in OS surfaces is treated differently than when PCC data is exposed to apps. Exposure in the OS is considered trusted and necessary for technical and practical reasons, such as needing to surface a sharing permission request to the user. 
However, when data is exposed to users in a UI outside PCC there are three options to consider:

\begin{enumerate}
\item \emph{Appearance in an OEM-rendered system UI.} These UIs usually include containers like the app launcher or notifications panel, and the UI shows data from PCC via System APIs with specific purposes.
\item \emph{Appearance in AOSP-rendered system UI.} This includes a growing set of surfaces like the “Privacy dashboard” where data can be safely shown to the user.  
\item \emph{Appearance in delegated UI.} This is a mechanism where the rendering is done by the PCC feature but shown in a different app, while restricting the target app from observing the content or user actions via the isolation and controls mechanisms previously described. 
\end{enumerate}

\subsection{Data sources}
There are several existing Android APIs that are used as data sources for PCC features. In general, users and apps can opt-out of data being shared over these APIs. Two examples of these data source APIs, for which the requirements are described in the Android Compatibility Definition Document (CDD), are described below.  

\subsubsection{Content Capture API}
\href{https://developer.android.com/reference/android/view/contentcapture/ContentCaptureManager}{Content Capture API} is a mechanism introduced in Android 11 to allow the OS and PCC features to use screen data (under user and application control). Usage of this data is very limited even within PCC, and never shared across apps without specific user actions. Features such as Smart Reply, described in detail later, rely on this API.\\

In Android, the standard user interface tools are instrumented so that data is automatically formatted and made available to the API when rendered on the screen. The API can also include other types of structured app data, specifically passed by the apps to PCC (i.e. share-data)\footnote{\url{https://developer.android.com/reference/android/view/contentcapture/ContentCaptureManager\#shareData(android.view.contentcapture.DataShareRequest,\%20java.util.concurrent.Executor,\%20android.view.contentcapture.DataShareWriteAdapter)}}. \\

Third-party applications can always opt-out of their data being passed through the API to PCC features, and this is automatically enforced by the operating system. Applications can also request that all instances of a specific piece of data (e.g. a message that was deleted by the user) be deleted by specifying its unique “locus id”\footnote{\url{https://developer.android.com/reference/android/content/LocusId}} generated by the API. 

\subsubsection{AppSearch}
\href{https://developer.android.com/develop/ui/views/search/appsearch}{AppSearch} is a general purpose indexing and searching mechanism, provided by the Android OS to all app developers\footnote{\url{https://developer.android.com/guide/topics/search/appsearch}}. This is designed to run optimally on-device and fulfill most applications needs. It was introduced in Android 12, with backwards compatibility via Android Jetpack\footnote{The backwards compatibility implementation via Android Jetpack means that prior to Android 12, data is not shared outside of the client app. 
}.\\

AppSearch also allows data to be shared with different services, and it is shared with PCC on devices where it is available, but application developers can opt-out of this.\\

%% file: 03_Implementation_examples.tex
\section{Implementation examples}
Following the PCC architecture previously described, Android System Intelligence (ASI) sandboxing is achieved by the system rules, for example \texttt{<allow-association>} system (Table 2), or the absence of \texttt{android.permission.INTERNET}\footnote{\url{https://developer.android.com/reference/android/Manifest.permission\#INTERNET}} permission (through the permission system). It solely relies on PCS and the Android framework (via different APIs) for any information egress.

\begin{table}[ht]
\tiny
\texttt{
\begin{itemize}
  \item[] \color{black}<allow-association \color{violet}target=\color{OliveGreen}"com.google.android.as"\color{violet} allowed=\color{OliveGreen}"com.android.bluetooth"\color{black} />
 \item[] <allow-association \color{violet}target=\color{OliveGreen}"com.google.android.as"             \color{violet}allowed=\color{OliveGreen}"com.android.providers.contacts" \color{black}/>
 \item[] <allow-association \color{violet}target=\color{OliveGreen}"com.google.android.as" \color{violet}allowed=\color{OliveGreen}"com.android.providers.media"\color{black} />
 \item[] <allow-association \color{violet}target=\color{OliveGreen}"com.google.android.as" \color{violet}allowed=\color{OliveGreen}"com.android.providers.telephony"\color{black} />
 \item[] <allow-association \color{violet}target=\color{OliveGreen}"com.google.android.as" \color{violet}allowed=\color{OliveGreen}"com.android.systemui"\color{black} />
 \item[] allow-association \color{violet}target=\color{OliveGreen}"com.google.android.as" \color{violet}allowed=\color{OliveGreen}"com.google.android.providers.media.module"\color{black} />
 \item[] // Private Compute Services
 \item[] allow-association \color{violet}target=\color{OliveGreen}"com.google.android.as" \color{violet}allowed=\color{OliveGreen}"com.google.android.as.oss"\color{black} />
\end{itemize}}
\normalsize
\caption{Allow-association configuration for ASI}
\end{table}

The \texttt{<allow-association>} mechanism blocks any incoming and outgoing inter-process requests through primitives like \texttt{bindService()}, using content providers or content resolvers, broadcasting intents and others.\\

In addition to the \texttt{<allow-association>} mechanism Android System Intelligence follows the Android permission model. The APK publicly requests the required permissions (Table 3) in its manifest, and are typically pre-granted by the OS (so ASI must pass both the association and any traditional permission checks). In general, it is considered that data access permissions (e.g. \texttt{android.permission.ACCESS\_WIFI\_STATE}) are safe to be granted to the APK given the technical limitations for data sharing imposed by PCC.\\

On Pixel devices ASI APK fulfills the following roles\footnote{\url{https://developer.android.com/reference/android/app/role/RoleManager}}, defined by the OS\footnote{\url{https://cs.android.com/android/platform/superproject/+/master:packages/modules/Permission/PermissionController/res/xml/roles.xml}}:
\scriptsize
\texttt{
\begin{itemize}
    \item []System UI Intelligence (SYSTEM\_UI\_INTELLIGENCE)
    \item []System Ambient Audio Intelligence (SYSTEM\_AMBIENT\_AUDIO\_INTELLIGENCE)
    \item []System Audio Intelligence (SYSTEM\_AUDIO\_INTELLIGENCE)
    \item []System Notification Intelligence (SYSTEM\_NOTIFICATION\_INTELLIGENCE)
    \item []System Text Intelligence (SYSTEM\_TEXT\_INTELLIGENCE)
    \item []System Visual Intelligence (SYSTEM\_VISUAL\_INTELLIGENCE)
\end{itemize}}

\begin{table}[ht]
\tiny
\color{OliveGreen}
\begin{verbatim}
  android.permission.ACCESS_BACKGROUND_LOCATION
  android.permission.ACCESS_COARSE_LOCATION
  android.permission.ACCESS_FINE_LOCATION
  android.permission.ACCESS_NETWORK_STATE
  android.permission.ACCESS_SHORTCUTS
  android.permission.ACCESS_WIFI_STATE
  android.permission.BLUETOOTH_ADMIN
  android.permission.BLUETOOTH_CONNECT
  android.permission.BLUETOOTH_SCAN
  android.permission.CAMERA
  android.permission.CAPTURE_AUDIO_HOTWORD
  android.permission.CAPTURE_AUDIO_OUTPUT
  android.permission.CAPTURE_MEDIA_OUTPUT
  android.permission.CAPTURE_VOICE_COMMUNICATION_OUTPUT
  android.permission.CONTROL_INCALL_EXPERIENCE
  android.permission.EXEMPT_FROM_AUDIO_RECORD_RESTRICTIONS
  android.permission.FOREGROUND_SERVICE
  android.permission.MANAGE_APP_PREDICTIONS
  android.permission.MANAGE_MUSIC_RECOGNITION
  android.permission.MANAGE_SEARCH_UI
  android.permission.MANAGE_SMARTSPACE
  android.permission.MANAGE_SOUND_TRIGGER
  android.permission.MANAGE_UI_TRANSLATION
  android.permission.MODIFY_AUDIO_ROUTING
  android.permission.MODIFY_AUDIO_SETTINGS
  android.permission.MODIFY_PHONE_STATE
  android.permission.OBSERVE_SENSOR_PRIVACY
  android.permission.PACKAGE_USAGE_STATS
  android.permission.QUERY_ALL_PACKAGES
  android.permission.READ_CALL_LOG
  android.permission.READ_CONTACTS
  android.permission.READ_DEVICE_CONFIG
  android.permission.READ_EXTERNAL_STORAGE
  android.permission.READ_GLOBAL_APP_SEARCH_DATA
  android.permission.READ_OEM_UNLOCK_STATE
  android.permission.READ_PEOPLE_DATA
  android.permission.READ_PHONE_STATE
  android.permission.READ_SMS
  android.permission.RECEIVE_BOOT_COMPLETED
  android.permission.RECORD_AUDIO
  android.permission.REQUEST_NOTIFICATION_ASSISTANT_SERVICE
  android.permission.START_ACTIVITIES_FROM_BACKGROUND
  android.permission.SUBSTITUTE_NOTIFICATION_APP_NAME
  android.permission.SYSTEM_APPLICATION_OVERLAY
  android.permission.SYSTEM_CAMERA
  android.permission.UNLIMITED_SHORTCUTS_API_CALLS
  android.permission.UPDATE_DEVICE_STATS
  android.permission.VIBRATE
  android.permission.WAKE_LOCK
  android.permission.WRITE_SECURE_SETTINGS
  com.android.alarm.permission.SET_ALARM
  com.google.android.ambientindication.permission.AMBIENT_INDICATION
  com.google.android.apps.nexuslauncher.permission.HOTSEAT_EDU
  com.google.android.setupwizard.SETUP_COMPAT_SERVICE
\end{verbatim}
\color{black}
\caption{Permissions requested by ASI APK, at the time of writing}
\end{table}
\normalsize

\newpage ASI is covered by the CDD for \texttt{ContentCaptureService}, \texttt{AppSearch} Services\footnote{\url{https://source.android.com/docs/compatibility/12/android-12-cdd\#986_content_capture_and_app_search}} and Android Intelligence roles\footnote{\url{https://source.android.com/compatibility/12/android-12-cdd\#91_permissions}: [C-4-1...C-4-3]
} as it has access to screen and app search content, and other data. This means that ASI follows CDD guidance such as:
\scriptsize
\texttt{
\begin{itemize}
 \item[] 9.8.6 [C-SR] Are STRONGLY RECOMMENDED NOT to request the INTERNET permission.
 \item[] 9.8.6 [C-SR] Are STRONGLY RECOMMENDED to only access the internet through structured APIs backed by publicly available open-source implementations.`
\item[] 9.8.6 [C-SR] Are STRONGLY RECOMMENDED to keep the services separate from other system components(e.g. not binding the service or sharing process IDs) except for the following: Telephony, Contacts, System UI, and Media
\end{itemize}}
\normalsize

These requirements are enforced (i.e. turned into MUST instead of STRONGLY RECOMMENDED) for packages holding the \texttt{SYSTEM\_*\_INTELLIGENCE} roles (mentioned above).\\

Below are some example features provided by Android System Intelligence as implemented in Pixel and potentially other devices for Android 12.
\newpage
\subsection{Smart Reply}
Smart Reply (part of Smart Input\cite{12}) suggests quick responses on the keyboard based on both previous and current screen content. To do this, ASI extracts relevant entities from different apps such as addresses, names, and others, and then offers them to the user as suggestions. These suggestions show up in compatible keyboards, but leverage privacy preserving approaches to avoid revealing the content to the keyboard systems.\\ 

\begin{figure*}[ht]
\centering
\includegraphics[width=0.95\textwidth]{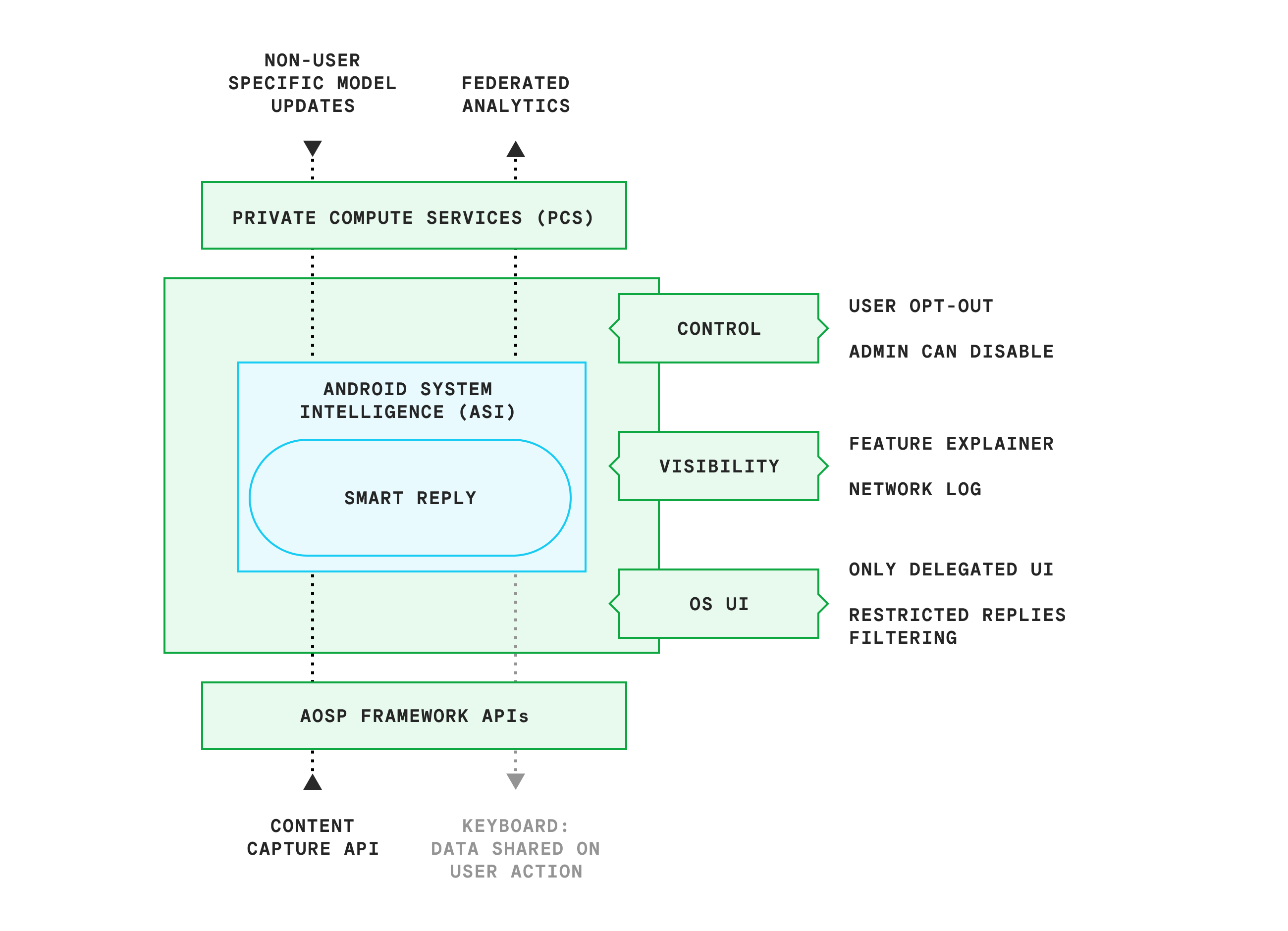}
\caption{Smart Reply Implementation}
\end{figure*}

This feature is implemented as follows:
\begin{itemize}
\item Uses Content Capture API as a data source, which is an Android Framework API with clear control in the framework. This ensures limits on access by the feature.
\item Users (via Settings) and app-developers (via \texttt{FLAG\_SECURE} or overriding the API) can opt-out of screen data being used by this feature.
\item Device administrators can disable screen capture for enterprise devices by using the \texttt{POLICY\_DISABLE\_SCREEN\_CAPTURE}\footnote{\url{https://developer.android.com/reference/android/app/admin/DevicePolicyManager\#POLICY\_DISABLE\_SCREEN\_CAPTURE}} policy or through Android Management\footnote{\url{https://developers.google.com/android/management/reference/rest/v1/enterprises.policies}} cloud API. This ensures that admins retain control of what data is accessible to features and services.
\item The functionality is explained\cite{12} and can be disabled by the user in the UI.\newpage
\item Users can specifically allow data to leave PCC
\begin{itemize}
\item To display the candidate responses to the user in a keyboard which supports it such as Gboard, a delegated UI via a specific Android Framework API is used, which means that no data leaves the PCC boundary at rendering time.
\item Candidate filtering by input is disabled after a series of keystrokes, due to the keyboard being able to obtain how many candidate responses are displayed. This prevents attempts to guess the content of the responses prior to the user tapping on the chip. 
\item Data is only held ephemerally in PCC (kept in memory with short TTLs\footnote{TTL or Time-To-Live refers to the time that data is kept before being deleted.}, thus suggestions are based on the data observed recently).
\item Data is only obtained from apps that PCC understands how to infer data from (by supplying an allowlist to the system).
\end{itemize}
\item Uses Private Compute Services’ OSS APIs for all network access
\begin{itemize}
\item ML models are non-user specific.
\item Analytics is performed through federated analytics with secure aggregation, with additional limitations to reduce risk (e.g. restricting the analysis to only popular apps, limited TTLs of data on-device, etc). Data semantics being used for federated analytics is documented in the open-source code in a form of policy that is enforced in the close-source code in PCC. 
\end{itemize}
\end{itemize}

\newpage
\subsection{Live Caption}
This feature provides captions for any content playing on the phone, using best-in-class on-device speech recognition. To do this, it processes all audio playing on the Android device and displays the transcript as an overlay which is an AOSP-rendered UI (data is not accessible to apps).

\begin{figure*}[ht]
\centering
\includegraphics[width=0.95\textwidth]{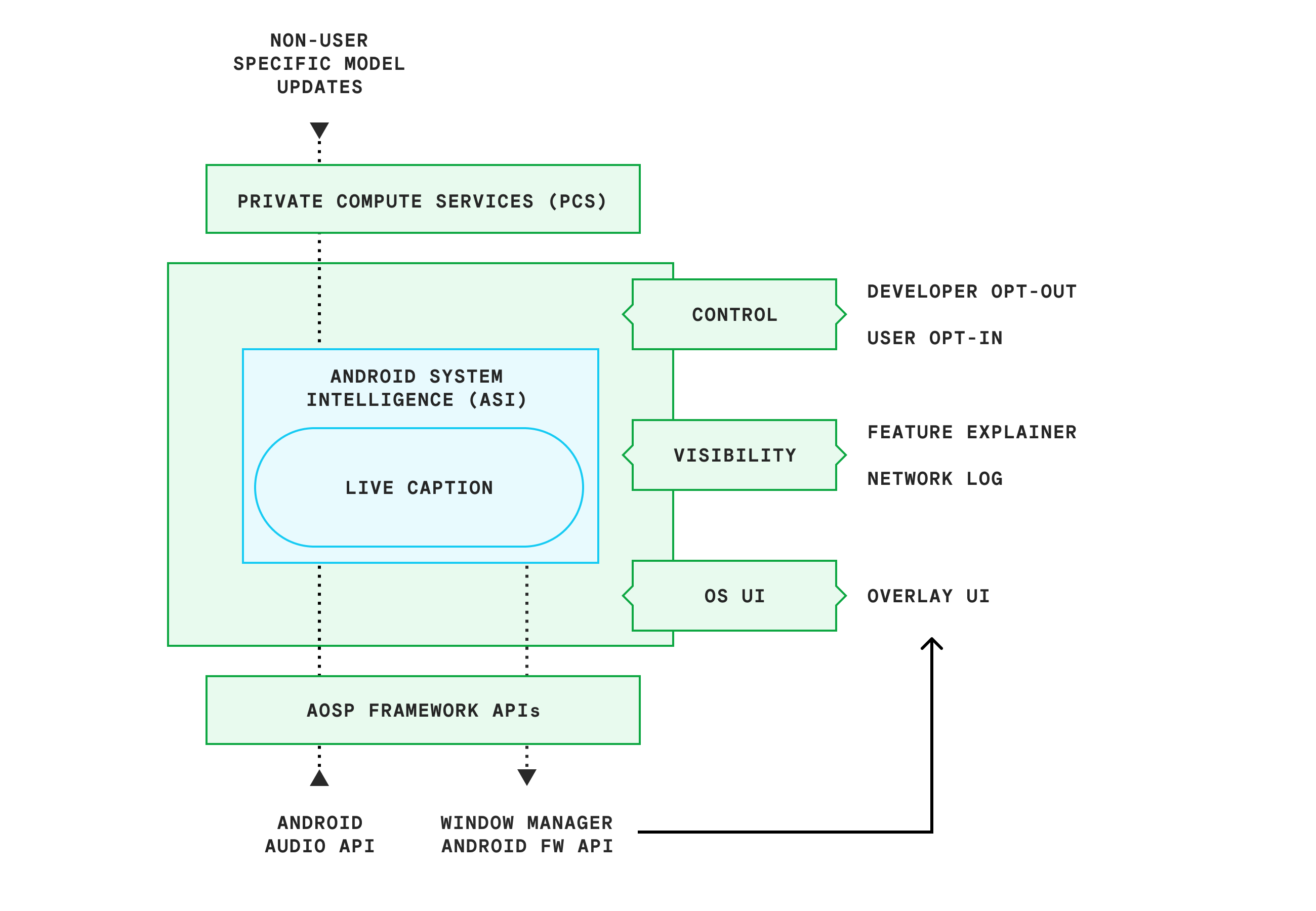}
\caption{Live Caption Implementation}
\end{figure*}

This feature is implemented as follows:
\begin{itemize}
\item Uses Android Audio APIs as a data source, which is an Android Framework API with clear control in the framework, which ensures limits on access by the feature. 
\begin{itemize}
\item App-developers (via \texttt{ALLOW\_CAPTURE\_BY\_NONE}) can opt-out of data being used by this feature. 
\end{itemize}
\item The feature is turned on by the user and shows a clear UI when on. The Live Caption \href{https://support.google.com/accessibility/android/answer/9350862}{settings screen} details how the feature works and what data is used. 
\item Data does not leave PCC and is only rendered in a system surface 
\begin{itemize}
\item To display the transcript the feature uses a Window Manager Android Framework API to create an overlay. \item Data is only held ephemerally in PCC (kept in memory with short TTLs).
\end{itemize}
\item Uses Private Compute Services’ OSS APIs for all network access
\begin{itemize}
\item Model updates are non-user specific. 
\end{itemize}
\end{itemize}
\newpage
\subsection{Screen attention}
When enabled, this feature prevents screen deactivation while the user is still looking at their phone. At the time screen dimming is scheduled to happen, if a face is detected in the front-facing camera's view, then the dimming is postponed.

\begin{figure*}[ht]
\centering
\includegraphics[width=0.95\textwidth]{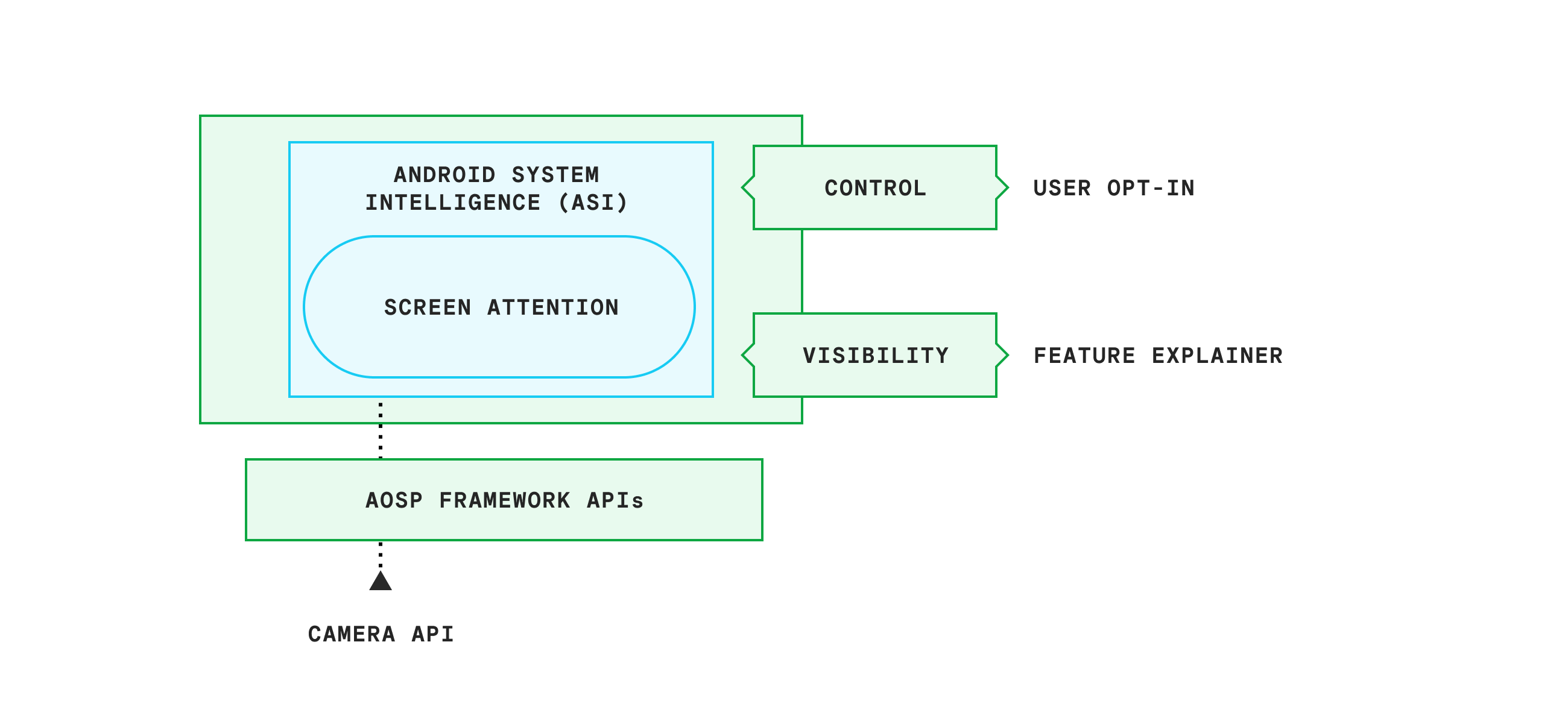}
\caption{Screen Attention Implementation}
\end{figure*}

This feature is implemented as follows:
\begin{itemize}
\item Uses standard Camera APIs as a data source, which is an Android Framework API with clear control in the framework. This ensures limits on access by the feature.  
\item The user can turn on this feature in a UI\footnote{\url{https://support.google.com/pixelphone/answer/6111557}} where it is clearly explained. The feature is not enabled without this explicit opt-in.
\item Data does not leave PCC and is only processed within PCC and the OS through the \texttt{AttentionManagerService} Framework API. Data is only held ephemerally (kept in memory with short TTLs).
\item Does not use any of the network capabilities, as the models are only updated via the APK.

\end{itemize}
\newpage

%% file: 04_conclusion.tex
\section{Conclusion}
Private Compute Core, as defined in Android 12, has introduced mechanisms that allow for the creation of high-value features that use ambient and OS-level data while following Android guidelines and privacy-preserving methodologies. Thanks to Private Compute Services, these features can still use cloud services in limited and privacy-preserving ways, through OSS that can be reviewed by external experts. Android System Intelligence, as available on all Pixel phones and other Android devices, represents the main vehicle for PCC features. As PCC grows, we expect to expand the space and include many new features, following and improving upon the mechanisms outlined in this paper.

\section{Acknowledgements}
The PCC project is building from the work of and relying on the support of Dianne Hackborn, who besides establishing much of the Android privacy and security model, created the kernel of this project and has been the strongest supporter and most engaged technical leader a project can ask for. Together with Dianne, Tom Hume, Wei Huang, and Asela Gunawardana have been instrumental leaders in getting PCC started and to the state it is today, as have our privacy and security team with Jordan McClead and Vishwath Mohan in particular. Finally the PCC project team would like to thank executive sponsors Dave Kleidermacher, Suzanne Frey, Matej Pfajfar, Shan Rao, Charmaine D'Silva, Sagar Kamdar, Andrei Popescu, Blaise Agüera y Arcas, David Petrou and the hundreds of engineers, product managers, user experience designers and project managers that have contributed to the guidelines, infrastructure and features in PCC and the Android framework code that supports it.

%% file: references.bib
@misc{1,
    author    = "Suzanne Frey",
    title     = "Introducing Android’s Private Compute Services",
    howpublished = "\url{https://security.googleblog.com/2021/09/introducing-androids-private-compute.html}",
    year     = "2021",
    month    = "09",
    note     = "Google Security Blog",
    }

@misc{2,
    author    = "Sameer Samat",
    title     = "Android 12 Beta: Designed for you",
    howpublished = "\url{https://blog.google/products/android/android-12-beta/}",
    year     = "2021",
    month    = "05",
    note     = "The Keyword Blog",
    }

@article{3,
    author    = "Latanya Sweeney",
    title     = "K-Anonymity: A Model for Protecting Privacy",
    journal   = "International Journal of Uncertainty, Fuzziness and Knowledge-Based Systems",
    volume   = "10",
    number   = "05",
    pages    = "557-570",
    year      = "2002",
    note     = "DOI: \url{https://doi.org/10.1142/S0218488502001648}",
    }

@misc{4,
    author    = "Michelle Tadmor-Ramanovich and Nadav Bar",
    title     = "On-Device Captioning with Live Caption",
    howpublished = "\url{https://ai.googleblog.com/2019/10/on-device-captioning-with-live-caption.html}",
    year     = "2019",
    month    = "10",
    note     = "Google AI Blog.",
    }

@misc{5,
    author    = "Beat Gfeller and Blaise Aguera-Arcas and Dominik Roblek and James David Lyon and Julian James Odell and Kevin Kilgour and Marvin Ritter and Matt Sharifi and Mihajlo Velimirović Ruiqi Guo and Sanjiv Kumar",
    title     = "Now Playing: Continuous low-power music recognition",
    howpublished = "\url{https://research.google/pubs/pub46522}",
    year     = "2017",
    month    = "10",
    note     = "NIPS 2017 Workshop: Machine Learning on the Phone",
    }

@misc{7,
    author    = "Brendan McMahan and Daniel Ramage",
    title     = "Federated Learning: Collaborative Machine Learning without Centralized Training Data",
    howpublished = "\url{https://ai.googleblog.com/2017/04/federated-learning-collaborative.html}",
    year     = "2017",
    month    = "04",
    note     = "Google AI Blog", 
    }

@misc{8,
    author    = "Daniel Ramage and Stefano Mazzocchi",
    title     = "Federated Analytics: Collaborative Data Science without Data Collection",
    howpublished = "\url{ https://ai.googleblog.com/2020/05/federated-analytics-collaborative-data.html}",
    year     = "2020",
    month    = "05",
    note     = "Google AI Blog",
    }

@misc{9,
    author    = "K. A. Bonawitz and Vladimir Ivanov and Ben Kreuter and Antonio Marcedone and H. Brendan McMahan and Sarvar Patel and. Daniel Ramage and Aaron Segal and Karn Seth",
    title     = "Practical Secure Aggregation for Federated Learning on User-Held Data",
    howpublished = "\url{https://research.google/pubs/pub45808/}",
    year     = "2016",
    note     = "NIPS Workshop on Private Multi-Party Machine Learning",
    }

@article{10,
	author		= "Benny Chor and Oded Goldreich and Eyal Kushilevitz and Madhu Sudan",
	title		= "Private information retrieval",
	journal	= "In Proceedings of IEEE 36th Annual Foundations of Computer Science",
    note     = "DOI: \url{https://doi.org/10.1109/SFCS.1995.492461}",
	address	= "Milwaukee, WI, USA",
	howpublished = "\url{https://ieeexplore.ieee.org/abstract/document/492461}",
	year		= "1995",
	month		= "10",
}

@misc{11,
    author    = "René Mayrhofer and Jeffrey Vander Stoep and Chad Brubaker and Nick Kralevich",
    title     = "The Android Platform Security Model",
    howpublished = "\url{https://arxiv.org/pdf/1904.05572.pdf}",
    year      = "2020",
}

@misc{12,
    author    = "Yang Lu and Angana Ghosh and Xu Liu",
    title     = "Privacy-Preserving Smart Input with Gboard",
    howpublished = "\url{https://security.googleblog.com/2020/10/privacy-preserving-smart-input-with.html}",
    year     = "2020",
    month    = "05",
    note     = "Google Security Blog", }
